# Longest paths in Planar DAGs in Unambiguous Logspace

Nutan Limaye and Meena Mahajan and Prajakta Nimbhorkar
The Institute of Mathematical Sciences, Chennai 600 113, India.
{nutan,meena,prajakta}@imsc.res.in

October 23, 2018

## Abstract

We show via two different algorithms that finding the length of the longest path in planar directed acyclic graph (DAG) is in unambiguous logspace UL, and also in the complement class co-UL. The result extends to toroidal DAGs as well.

## 1 Introduction

Consider the following problems in graphs:

$$
\begin{aligned}
\text{Reach} &= \{ (G,s,t) \mid G \text{ contains a path from } s \text{ to } t \} \\
\text{Distance} &= \{ (G,s,t,k) \mid G \text{ contains a path of length } \leq k \text{ from } s \text{ to } t \} \\
\text{Long-Path} &= \{ (G,s,t,k) \mid G \text{ has a simple path of length } \geq k \text{ from } s \text{ to } t \}
\end{aligned}
$$

These problems have widely differing complexities: some of the results below are folklore, some are recent advances. Reach is NL-complete for general graphs and remains NL-hard even if the graphs are acyclic. It is L-complete for undirected graphs [Rei05], and is sandwiched between L and UL ∩ co-UL for planar directed graphs [BTV07]. Distance is NL-complete for general graphs, and remains NL-hard even if the graphs are acyclic, or if the graphs are undirected, but it is in UL ∩ co-UL for planar directed graphs [TW07]. Long-Path is NP-complete for general graphs, since it includes Hamiltonian paths as a special case. It remains NP-hard for planar undirected graphs. It is NL-complete for directed acyclic graphs. However its complexity for planar directed acyclic graphs is, to the best of our knowledge, not yet studied.

In this note we consider this combination of planarity and acyclicity for Long-Path. Our main result is:

**Theorem 1** PDLP, *the* Long-Path *problem for planar directed acyclic graphs, is in* UL ∩ co-UL.

Thus Long-Path shares the current best-known upper bounds for Reach and Distance for such graphs. We also address the question of when the three problems are indeed equivalent on DAGs, and give partial bounds (Theorem 8,10). A recent result in [JT07] shows that for an important subclass of planar DAGs, namely series-parallel graphs, the three problems are indeed equivalent and are all L-complete. Theorem 1 is in fact an unobserved corollary of their construction (see also [JT06]. An analogous result for planar DAGs equating the three problems would be nice, but is not known.

For graphs with embeddings on the torus, [ADR05] shows that reachability is no harder than planar reachability. We observe that Distance and Long-Path are also no harder than the planar versions (Corollary 6).

## 2 Known results, and Preprocessing

We use the following results:

**Lemma 2 ([BTV07])** Reach *in planar directed graphs is in* UL ∩ co-UL.

**Lemma 3 ([TW07])** Distance *in planar directed graphs is in* UL ∩ co-UL.

**Lemma 4 ([JT07])** Distance *and* Long-Path *in series-parallel graphs are equivalent.*

**Lemma 5 ([ADR05])** Reach*(Torus) logspace-many-one reduces to planar* Reach.

For any subclass $C$ of graphs, let Reach($C$), Distance($C$), and Long-Path($C$) denote the restriction of these problems to instances from $C$.

For directed acyclic graphs, $(G, s, t) \in$ Reach $\Leftrightarrow (G, s, t, |V|) \in$ Distance $\Leftrightarrow (G, s, t, 0) \in$ Long-Path. So Distance($C$) and Long-Path($C$) are at least as hard as Reach($C$) for any subclass $C$ of directed acyclic graphs.

Consider any directed acyclic instance $(G, s, t, k)$ of Distance or Long-Path. By (parallel) queries of the form $(G, s, u)$ or $(G, u, t)$ to Reach, we can remove all vertices that do not figure on some $s$-to-$t$ path to obtain in $\mathsf{L}^{\mathsf{Reach}}$ a single-source ($s$) single-sink ($t$) graph $G'$, and all queries to Reach involve only the graph $G$. So now onwards we only consider the case where we want to find a long path between the unique source and the unique sink.

If the input graph $G$ is not planar but can be embedded on a torus, then we use the construction of Lemma 5. This gives a planar graph $G'$ with the following properties: There are $l \in O(n)$ copies of $G$ cut and stitched together, and hence there are $l$ vertices $t_1, \ldots, t_l$ and one special vertex, say $s_1$, such that

$$\exists \rho : s \to_G^* t \wedge |\rho| = l \Leftrightarrow \exists i \exists \rho_i : s_1 \to_{G'}^* t_i \wedge |\rho_i| = l$$

Hence

**Corollary 6**
$$\mathsf{Distance}(Torus) \leq_m^{\log} \mathsf{Distance}(Planar)$$
$$\mathsf{Long\text{-}Path}(Toroidal\ DAGs) \leq_m^{\log} \mathsf{Long\text{-}Path}(Planar\ DAGs)$$

## 3 Algorithm using distance computation

Our first algorithm for Long-Path in planar DAGs uses a simple extension of Lemma 4.

**Lemma 7** Distance *and* Long-Path *in planar directed acyclic graphs are equivalent modulo planar reachability.*

This actually follows from [JT07] itself; though they claim their result only for series-parallel graphs, it works for single-source single-sink acyclic graphs as well, where $s$ and $t$ in the input instance are the source and sink respectively. It doesn't even seem to use planarity. To make this clear, we present below in Theorem 8 their proof simplified by specialising to unweighted graphs, and stated with minimum conditions. Lemma 7 is an obvious corollary.

**Theorem 8** *Let $C$ be any subclass of directed acyclic graphs. There is a function $f$, computable in* L *with oracle access to* Reach$(C)$*, that reduces* Distance$(C)$ *to* Long-Path$(C)$ *and* Long-Path$(C)$ *to* Distance$(C)$.

**Proof:** Let $G = (V, E)$ be a directed acyclic graph with a unique source $s$ and a unique sink $t$. Every vertex of $G$ lies on some $s \to^* t$ path. ($G$ is unweighted, so all edges have weight 1.) Let $M$ be the number of edges in $G$. Construct a new graph $G' = (V', E')$ as follows:

For each $u \in V$, define $P_u = \{x \in V \mid x \to_G^* u\}$. Since $s$ is the unique source, $\forall u, s \in P_u$. Also define $E_u = \{\langle x, y \rangle \mid x \in P_u, y \notin P_u\}$. Since $G$ is acyclic, $\forall \langle x, y \rangle \in E, \langle x, y \rangle \in E_x$.

Let $\rho$ be any $s \to^* t$ path. For every vertex $u \in V$, $|\rho \cap E_u| = 1$. Why? Note that $s \in P_u$, $t \notin P_u$, and along the path $\rho$, we transit from being in $P_u$ to being outside $P_u$ exactly once. Let this transition occur on edge $\langle x, y \rangle$. Then $\langle x, y \rangle \in E_u$, and no other edge of $\rho$ can be in $E_u$.

To obtain $G'$, we replace each edge $e = \langle u, v \rangle$ by a path of length $l_{uv}$ determined as follows:

$$l_{uv} = 2\left(\sum_{x \in V : e \in E_x} \text{out-degree}(x)\right) - 1 = 2\left(\sum_{x \in V : u \in P_x, v \notin P_x} \text{out-degree}(x)\right) - 1$$

Since $G$ is acyclic, the vertex $u$ itself always qualifies in the above sum, and so $l_{uv}$ is positive.

Now the crucial claim: each $s \to^* t$ path $\rho$ in $G$, of length $|\rho|$ in terms of number of edges, is transformed by the above to a path in $G'$ of length *exactly* $2|E| - |\rho|$. This is because the length of the transformed path is

$$\begin{aligned}
\sum_{uv \in \rho} l_{uv} &= \sum_{uv \in \rho}\left[2\left(\sum_{x \in V : u \in P_x, v \notin P_x} \text{out-degree}(x)\right) - 1\right] \\
&= 2\left(\sum_{uv \in \rho} \sum_{x \in V : u \in P_x, v \notin P_x} \text{out-degree}(x)\right) - |\rho| \\
&= 2\sum_{x \in V}\left(\text{out-degree}(x) \sum_{e \in \rho \cap E_x} 1\right) - |\rho| \\
&= 2\left(\sum_{x \in V} \text{out-degree}(x) \cdot |\rho \cap E_x|\right) - |\rho| \\
&= 2\sum_{x \in V} \text{out-degree}(x) - |\rho| = 2|E| - |\rho|
\end{aligned}$$

It thus follows that the longest (shortest) path in $G$ is mapped to the shortest (longest, respectively) path in $G'$. In fact, if the $s \to^* t$ paths are ordered monotonically with respect to length, then the above transformation precisely reverses this ordering. Hence the reduction function $f$ maps $(G, s, t, k)$ to $(G', s, t, 2|E| - k)$.

The next crucial observation: $G'$ can be obtained from $G$ in logspace with oracle access to Reach, where all queries involve only the graph $G$. This is because obtaining $G'$ merely involves finding the sets $P_u, E_u$. ∎

Theorem 1 follows from Theorem 8 and Lemma 3.

## 4 Algorithm using inductive counting

There is another method to obtain Theorem 1, bypassing Theorem 8 but using Lemmas 2,3. We sketch it here because it is instructive to see how double inductive counting can be used, and also because it says something more general as well: it places Long-Path in (UL ∩ co-UL)⊕Reach(C) for any family $C$ of acyclic max-unique graphs (Theorem 10).

The initial steps are similar to those used in [TW07] to place planar Distance in UL ∩ co-UL.

1. Given a graph $\hat{G}$, make it single-source single-sink $G$ as described in the preprocessing step. Reduce the degree of each vertex to 3. (To reduce the degree of nodes, in [ADR05] a vertex of degree $d$ is replaced by a cycle of length $d$. Since we cannot afford to introduce cycles, we use the trick of [CD06]; insert incoming and outgoing trees at each vertex.) This construction maps edges to paths, and we can identify a unique new edge as "responsible" for each original edge. We mark such edges.

   Embed $G$ into a grid using the [ADR05] reachability-preserving construction. The output of this step is a grid graph $G'$, with the edges of $G$ (original edges) marked in $G'$ and is obtained in logspace. If the original graph $G$ had $n$ vertices, the new grid graph is of dimensions $n^2 \times n^2$.

2. The graph $G'$ is then subject to a weighting scheme building upon that of [BTV07], and can be described as follows: every horizontal edge $e$ gets weight $n^4 + (\text{mark}(e) \times n^8)$, and every vertical edge $e$ gets weight $n^4 + (\text{mark}(e) \times n^8) + (\text{up}(e) \times \text{col}(e))$, where mark($e$) is one if the edge $e$ is marked; zero otherwise, col($e$) equals the column number in which the edge $e$ appears, and up($e$) is $+1$ if the edge $e$ is upwards, $-1$ otherwise. This is the graph $G''$.

3. The last step in [TW07] is to use the double counting technique of [RA97] on the min-unique graph $G''$. The idea here is to use the inductive counting counter $c_k$ that keeps track of number of vertices within distance $k$, and to use a cumulative paths counter $s_k$ that keeps track of the shortest paths of the nodes so counted. The first counter allows checking the complement of reachability, the second allows doing so unambiguously. As mentioned in [TW07], a third counter $m_k$ tracking cumulative marked edges can be added, allowing distance computation uniquely.

Can we directly use this strategy for long paths as well?

The argument of [TW07] concerning Step 2 is restricted to shortest paths; however, one can observe something more general about the above weighting scheme.

**Observation 9** *For any length $l$, all the st paths of length $l$ in $G$ will be mapped to paths of weight greater than $(l \times n^8)$ and less than $((l+1) \times n^8)$ in $G''$, and the maximum weight and the minimum weight paths in this range will be unique. Thus $G''$ is both* min-unique *and* max-unique: *for each pair $u, v$, if there is a path from $u$ to $v$, then the shortest and the longest paths are unique.*

Observation 9 already guarantees a max-unique graph.

Step 3 above can not be used as it is. For computing the shortest path, we can initialise $c_0 = 1$ and $\Sigma_0 = 0$. If the same semantics is to be used for computing the longest path, then $c_0$ should be the number of vertices having length of the longest path from $s$ at least 0, and should be initialised to $n$. However $\Sigma_0$ should then contain the total lengths of all the longest paths, which is an unknown quantity. To handle this, we redefine $\Sigma_k$ to be sum of lengths of the longest paths for those vertices whose longest path to $t$ is of length at most $k$. This allows a procedure similar to [RA97] to work correctly, but now it is no longer unambiguous. To make it unambiguous, we introduce more nondeterminism into the [RA97] procedure. We guess the sum of lengths of all the $u \to^* t$ longest paths *a priori* and tally it in the end with the final $s_k$.

The detailed procedures are given below, which imply the following:

**Theorem 10** *Let $G$ be a directed acyclic graph with a unique source $s$, a unique sink $t$, such that $G$ is max-unique. (For each pair $u, v$, if there is a path from $u$ to $v$, then the longest $uv$ path is unique.) Then the length of the longest $st$ path can be computed in* UL ∩ co-UL.

The proof follows from Claims 11, 12, 13 and 14.

Notation: $D(v) = $ Length of the longest path from $v$ to $t$.

$$S_k = \{v | D(v) \geq k\}, \quad c_k = |S_k|, \quad \Sigma_k = \sum_{v \in V \setminus S_k} D(v), \quad T = \sum_{v \in V} D(v)$$

---

**Algorithm 1** Main

   Input: $G, s, t$
   Guess nondeterministically $M = \sum_{v \in V} D(v)$. $n \leq M \leq n^2$.
   $c_0 \leftarrow n, \Sigma_0 \leftarrow 0, k \leftarrow 0$
   **while** $c_k \neq 0$ **do**
      $k \leftarrow k + 1$
      Update ($c_k$ and $\Sigma_k$)
   **end while**
   **if** $\Sigma_k \neq M$ **then**
      Halt and reject
   **else**
      Accept
   **end if**

---

**Algorithm 2** Update: Procedure for updating $c_k$ and $\Sigma_k$

   Input: $G, s, t, c_{k-1}, \Sigma_{k-1}$
   $c_k \leftarrow c_{k-1}, \Sigma_k \leftarrow \Sigma_{k-1}$
   **for all** $v \in V$ **do**
      **if** Test$(G, k-1, c_{k-1}, \Sigma_{k-1}, v)$=true **then**
         **if** for all out-neighbours $x$ of $v$, Test$(G, k-1, c_{k-1}, \Sigma_{k-1}, x)$=false **then**
            $c_k \leftarrow c_k - 1, \Sigma_k \leftarrow \Sigma_k + k - 1$
         **end if**
      **end if**
   **end for**

---

**Claim 11** *If the guessed value of $M$ is correct (i.e. $M = T$), then algorithm* Test, *given the correct values of $c_k$ and $\Sigma_k$ as input, reports a decision on exactly one path.*

**Algorithm 3** Test: An unambiguous procedure to determine if $D(v) \geq k$
---
Input: $G, k, c_k, \Sigma_k, v$
$count = n, sum = 0, path.to.v = true, sum' = 0$
**for all** $x \in V$ **do**
  Guess nondeterministically if $D(x) \geq k$
  **if** Guess is no **then**
    Guess a path of length $l < k$ from $x$ to $t$. {If this fails then reject and halt.}
    $count \leftarrow count - 1$
    $sum \leftarrow sum + l$
    **if** $x = v$ **then**
      $path.to.v =$false
    **end if**
  **else**
    Guess a path of length $l' \geq k$ from $x$ to $t$. {If this fails then reject and halt.}
    $sum' \leftarrow sum' + l'$
  **end if**
**end for**
**if** $count = c_k$ and $sum = \Sigma_k$ and $sum' + sum = M$ **then**
  return path.to.v
**else**
  Reject and halt.
**end if**

---

**Proof:** The procedure Test, on each run $R$, guesses an $x \rightarrow^* t$ path $R_x$ for each vertex $x$. Depending on its guess for $D(x) \geq k$, it adds the length of $R_x$ to either sum or sum'. Finally these have to add up to $M$ for Test to report a decision.

When $M = T$, $M$ is indeed the sum of all $D(x)$. This can match sum+sum' exactly when all the guessed paths $R_x$ are longest. Since $G$ is max-unique, this happens on exactly one run. ∎

**Claim 12** *For any guessed value of $M$, given the correct values of $c_k$ and $\Sigma_k$ as input, all paths of algorithm* Test *that do not lead to rejection always return the correct decision.*

**Proof:** As described in the preceding proof, each run of Test guesses a path $R_x$ for each $x$. It may guess a path of length shorter than $D(x)$, but not longer. Since count is decremented only when it guesses that $D(x) < k$, and for other guesses some witnessing path of length at least $k$ is found, at the end the value of count is at most as large as $c_k$.

Suppose on some run Test returns a decision. Then on this run count $= c_k$. Suppose further that the decision is wrong.
Case 1: $D(v) < k$, but Test reports that it is larger. This cannot happen, since Test has to find a witnessing path of length at least $k$.
Case 2: $D(v) \geq k$, but Test reports that it is smaller. Then this run of Test does not account for $v$ in count. So at the end of the run, count $< c_k$, a contradiction. ∎

**Claim 13** *If the queries $(D(v) \geq k)$ are answered correctly by* Test, *then given $c_{k-1}$ and $\Sigma_{k-1}$, the values of $c_k$ and $\Sigma_k$ are updated correctly by algorithm* Update.

**Proof:** Update starts by assuming that $S_k = S_{k-1}$ and so $c_k = c_{k-1}$. Note that $S_k \subseteq S_{k-1}$, so Update only has to detect when to remove vertices from its current $S_k$.

For each $v$, Update checks whether $D(v) \geq k-1$ and $D(u) < k-1$ for all out-neighbours $u$ of $v$. If this holds, then the longest path from $v$ to $t$ is of length exactly $k-1$ and $v \notin S_k$. Thus the procedure decrements $c_k$ by 1 and increments $\Sigma_k$ by $k-1$.

So if all the queries are answered correctly by Test, then what Update does is correct. ∎

**Claim 14** *The algorithm* Main *is correct and unambiguous.*

**Proof:** Main starts with the correct values of $c_0$ and $\Sigma_0$. From claims 12 and 13, the correctness of Main is immediate. In particular, the final value of $\Sigma_k$ is always correct.

If $M = T$, then by Claim 11, procedure Test always returns a decision, unambiguously. Thus exactly one path of Main (amongst those where $M = T$ was guessed) leads to a decision, and this decision is correct.

If $M > T$, then no run of Test, at any stage $k$, can trace paths adding up to $M$. So Test, and hence Update, and Main have no accepting run.

If $M < T$, consider the runs on which Test and Update proceed to finally compute $\Sigma_k$. Since Main is correct, we know that $\Sigma_k = T$. Now the check $M = \Sigma_k$ fails and Main rejects and halts. ∎

# References

[ADR05] Eric Allender, Samir Datta, and Sambuddha Roy. The directed planar reachability problem. In *Proc. 25th annual Conference on Foundations of Software Technology and Theoretical Computer Science (FSTTCS)*, pages 238–249., 2005.

[BTV07] Chris Bourke, Raghunath Tewari, and N V Vinodchandran. Directed planar reachability is in unambiguous logspace. In *to appear in Proceedings of IEEE Conference on Computational Complexity CCC*, pages –, 2007.

[CD06] Tanmoy Chakraborty and Samir Datta. One-input-face MPCVP is hard for L, but in LogDCFL. In *Proc. of 26th FST TCS Conference, LNCS vol. 4337*, pages 57–68, 2006.

[JT06] Andreas Jakoby and Till Tantau. Computing shortest paths in series-parallel graphs in logarithmic space. In *Complexity of Boolean Functions*, Dagstuhl Seminar Proceedings, 2006. http://drops.dagstuhl.de/opus/volltexte/2006/618.

[JT07] Andreas Jakoby and Till Tantau. Logspace algorithms for computing shortest and longest paths in series-parallel graphs. In *FSTTCS*, page to appear, 2007. see also [JT06].

[RA97] Klaus Reinhardt and Eric Allender. Making nondeterminism unambiguous. In *IEEE Symposium on Foundations of Computer Science*, pages 244–253, 1997.

[Rei05] Omer Reingold. Undirected *st*-connectivity in logspace. In *Proc. 37th STOC*, pages 376–385, 2005.

[TW07] Thomas Thierauf and Fabian Wagner. The isomorphism problem for planar 3-connected graphs is in unambiguous logspace. Technical Report TR07-068, ECCC, 2007. to appear in STACS 2008.